# Anisotropy of Dirac cones and van Hove singularity in an organic Dirac fermion system

Ayaka Mori, Mitsuyuki Sato, Takeshi Yajima, Takako Konoike[#], Kazuhito Uchida, and Toshihito Osada[*]

*Institute for Solid State Physics, University of Tokyo,*

*5-1-5 Kashiwanoha, Kashiwa, Chiba 277-8581, Japan.*

We propose an experimental method to examine the in-plane anisotropy of electronic structure in layered conductors. In the method, we measure the interlayer magnetoresistance as a function of in-plane magnetic field orientation. We applied it to an organic Dirac fermion system $\alpha$-(BEDT-TTF)$_2$I$_3$ to experimentally determine the orientation of the anisotropic Dirac cones. It is concluded that the long axis of the elliptic constant-energy contours of the Dirac cone is tilted by approximately $-30°$ from the crystalline **a**-axis to **b**-axis under hydrostatic pressures. Additionally, we observed a signature of van Hove singularity (which is a saddle point of the band dispersion) at 30–40 K above or below the Dirac point. The ridgeline of the saddle point is estimated as almost parallel to the crystalline **b**-axis.

I. INTRODUCTION

A layered organic conductor $\alpha$-(BEDT-TTF)$_2$I$_3$ where BEDT-TTF denotes bis(ethylenedithio)-tetrathiafulvalene has attracted significant attention due to its two-dimensional (2D) massless Dirac fermion state under high pressure, which is similar to that of graphene [1]. Coupling between BEDT-TTF conducting layers is extremely low (i.e., the interlayer transfer energy, $t_c$, is significantly lower than 1 meV), and thus the compound is typically considered as a 2D system. At ambient pressure, $\alpha$-(BEDT-TTF)$_2$I$_3$ exhibits a phase transition into the insulating phase due to charge ordering (CO) at $T_{CO} = 135$ K. The CO transition temperature decreases with increasing pressure. Above the critical pressure $P_c > 1.2$ GPa, the CO phase vanishes, and the metallic phase survives at low temperatures [2]. Based on the tight-binding band calculation in the metallic phase, each BEDT-TTF layer exhibits 2D band dispersion in which the conduction and valence bands contact at two points ($\mathbf{k}_0$ and $-\mathbf{k}_0$) to form a pair of Dirac cones (termed as valleys) [3, 4]. The Fermi level is fixed at the Dirac point due to crystal stoichiometry, and thus the system is considered as a 2D Dirac semimetal (SM). In contrast to graphene, the Dirac cones are tilted and anisotropic. Dirac points are located at general points in the 2D Brillouin zone and not at symmetric points. Although the band structure of $\alpha$-(BEDT-TTF)$_2$I$_3$ is discussed by the tight-binding approach [3-9] and the first principles calculation [10, 11], most of which are performed for uniaxial pressure condition, the Dirac SM state under hydrostatic pressure is not necessarily reproduced well.

The realization of a Dirac SM in $\alpha$-(BEDT-TTF)$_2$I$_3$ is experimentally suggested in indirect ways via negative interlayer magnetoresistance [12, 13], interlayer Hall effect [14, 15], temperature dependence of magnetoresistance [16], specific heat [17], thermoelectric power [18], and site-selective NMR measurements [19, 20]. Specifically, Ref. 16 reported that the thermal excitation from the ground Landau level (zero mode) to the first Landau level exhibits square-root field dependence, and this is characteristic of the massless Dirac fermions. The excitation energy is less than 2 meV below 10 T. Additionally, Ref. 20 indicated Dirac cone narrowing due to the interaction as well as the existence of van Hove singularity (which was predicted to be located at approximately 12 meV from the Dirac points) [19].

Conversely, an extant study suggested the coexistence of massive-hole-pockets with Dirac cones in the complete pressure range based on magnetotransport and thermopower measurements [21]. This two-carrier scenario is not consistent with the aforementioned experiments. Unfortunately, it is essentially difficult to directly determine the band dispersion by angle-resolved photoemission spectroscopy (ARPES) or scanning tunneling spectroscopy (STS), which are impossible in a pressure cell. In the present study, we do not adopt the two-carrier picture and assume that only tilted Dirac cones exist around Fermi energy.

Recently, an electronic structure is experimentally clarified in the "weak CO" state, which denotes the CO state just below the critical pressure, in a manner similar to the Dirac SM state. The weak CO state is a massive Dirac fermion state where a small gap opens in the Dirac cone [22]. In the weak CO state, anomalous behaviors are observed. For example, the spin gap remains finite although the transport gap vanishes [23]. The edge transport along the boundary between CO domains was discussed [24]. Furthermore, the possibility of a topological phase was considered in the weak CO state

[25].

The purpose of the study involves determining the anisotropy of Dirac cones by an experiment independent of any band models. Hence, we developed a magnetotransport method that investigates the dependence of interlayer resistance on in-plane magnetic field orientation as shown in Fig. 1(a). In the study, we first discuss the principle of the experimental method. Subsequently, we present the experimental results for α-(BEDT-TTF)$_2$I$_3$. Finally, we discuss the anisotropy of the Dirac cone and van Hove singularity in the Dirac SM and weak CO states.

## II. INTERLAYER RESISTANCE IN LAYERED CONDUCTORS UNDER IN-PLANE MAGNETIC FIELDS

We consider the interlayer transport in general layered conductors with weak interlayer coupling under an in-plane magnetic field $\mathbf{B} = (B_x, B_y, 0)$. This denotes the quantum mechanical generalization of the preceding argument based on semi-classical Boltzmann transport theory [26]. The effective Hamiltonian of the system with the Landau gage $\mathbf{A} = (B_y z, -B_x z, 0)$ is expressed as follows:

$$\widehat{H} = \epsilon\left(-i\frac{\partial}{\partial x} + \frac{e}{\hbar}B_y z, -i\frac{\partial}{\partial y} - \frac{e}{\hbar}B_x z\right) - 2t_c \cos\left(-ic\frac{\partial}{\partial z}\right),$$

(1)

where $\epsilon(k_x, k_y)$ is the energy dispersion of each 2D layer parallel to the $xy$-plane, and $t_c$ and $c$ denote the interlayer transfer energy and interlayer spacing, respectively. We consider the interlayer coupling $\widehat{H}' \equiv -2t_c \cos(-ic\partial/\partial z)$ as a perturbation. The energy and envelope function of unperturbed electronic states are given as follows:

$$E_{\mathbf{k},z_i} = \epsilon(k_x, k_y) \equiv \epsilon(\mathbf{k}),$$

(2)

$$F_{\mathbf{k},z_i}(\mathbf{r}) = \langle \mathbf{r}|\mathbf{k}, z_i\rangle = \frac{1}{\sqrt{S}} \exp\left\{i\left(k_x - \frac{e}{\hbar}B_y z_i\right)x + i\left(k_y + \frac{e}{\hbar}B_x z_i\right)y\right\}.$$

(3)

Here, $\mathbf{k}$, $z_i$, and $S$ denote the in-plane wave number $(k_x, k_y, 0)$, the $z$-coordinate of each layer, and system area, respectively. It should be noted that the crystal momentum $\hbar\mathbf{k}$ no longer corresponds to the canonical momentum. It is defined on each layer and not conserved on the occasion of interlayer hopping. Specifically, the perturbation matrix elements are given as follows:

$$\langle \mathbf{k}', z_i'|\hat{H}'|\mathbf{k}, z_i\rangle = -t_c\left(\delta_{\mathbf{k}',\mathbf{k}+\mathbf{Q}}\delta_{z_i',z_i+c} + \delta_{\mathbf{k}',\mathbf{k}-\mathbf{Q}}\delta_{z_i',z_i-c}\right).$$

(4)

Here, $\mathbf{Q} \equiv (-e/\hbar)\mathbf{c} \times \mathbf{B} = (ecB_y/\hbar, -ecB_x/\hbar, 0)$ denotes the shift of the wave number after the single tunneling process. It originates from the Aharonov–Bohm phase corresponding to the magnetic flux surrounded by the loop over two neighboring layers as shown in Fig. 1(b). When wave functions with wave numbers $\mathbf{k}$ and $\mathbf{k}'$ on two neighboring layers are coupled by interlayer tunneling, the phase of wave functions changes by $kL - k'L$ after making a circuit of the loop $L \times c$, where $L$ can be arbitrarily selected. The phase must correspond to the Aharonov–Bohm phase $(-e/\hbar)\oint \mathbf{A} \cdot d\mathbf{l} = (-e/\hbar)BLc$, and thus the wave number shift $\mathbf{k}' - \mathbf{k} = \mathbf{Q}$ is obtained.

Based on the tunneling picture for the interlayer transport in layered conductors, the lowest order contribution of the interlayer coupling to the complex interlayer conductivity $\tilde{\sigma}_{zz}(\omega)$ corresponds to the single tunneling process between two neighboring layers [27]. It is given by the Kubo formula as follows [12, 14, 27]:

$$\tilde{\sigma}_{zz}(\omega) = -\frac{2i\hbar}{(2\pi)^2 c}\left(\frac{et_c c}{\hbar}\right)^2$$

$$\times \sum_{\pm}\iint \frac{f(\epsilon(\mathbf{k})) - f(\epsilon(\mathbf{k}\pm\mathbf{Q}))}{\epsilon(\mathbf{k}\pm\mathbf{Q}) - \epsilon(\mathbf{k})}\frac{1}{\epsilon(\mathbf{k}\pm\mathbf{Q}) - \epsilon(\mathbf{k}) - \hbar\omega - \frac{i\hbar}{\tau_\mathbf{k}}}dk_x dk_y,$$

(5)

where $\tau_\mathbf{k}$ indicates the scattering relaxation time. We take the limit of weak magnetic field and derive the formula for DC conductivity as follows:

$$\sigma_{zz} = \frac{2}{\pi^2 c}\left(\frac{et_c c}{\hbar}\right)^2 \iint \left(-\frac{df(\epsilon(\mathbf{k}))}{d\epsilon(\mathbf{k})}\right)\frac{\tau_\mathbf{k}}{1 + \left\{\left(-\frac{e}{\hbar}\mathbf{v}(\mathbf{k})\times\mathbf{B}\right)\cdot\mathbf{c}\right\}^2 \tau_\mathbf{k}^2}dk_x dk_y.$$

(6)

Here, $\mathbf{v}(\mathbf{k}) \equiv (1/\hbar)(d\epsilon(\mathbf{k})/d\mathbf{k})$ denotes group velocity. The interlayer conductivity denotes the summation of the contributions from every $\mathbf{k}$-point with the weight $-df(\epsilon(\mathbf{k}))/d\epsilon(\mathbf{k})$. We assume a constant relaxation time, and thus the contribution corresponds to a maximum when the Lorentz force $(-e)\mathbf{v}(\mathbf{k})\times\mathbf{B}$ corresponds to zero, in other words, when the group velocity $\mathbf{v}(\mathbf{k})$ is parallel to the magnetic field. Therefore, with respect to a constant-energy contour with sufficient thermal distribution, the segment perpendicular to the magnetic field significantly contributes to interlayer conductivity. This is the general result for multilayer conductors and even applies in the case wherein interlayer coupling is incoherent.

In the section below, we apply the aforementioned model to a layered conductor where each layer exhibits tilted Dirac cone dispersion as schematically shown in Fig. 1(c). At low temperatures, the electrons and holes are thermally distributed around the apex of the tilted cones (Dirac points). The constant-energy contours around the Dirac point are almost elliptic in the 2D **k**-space. When the magnetic field is perpendicular to the long axis of the ellipses, the segments of constant-energy contours almost normal to the magnetic field correspond to the longest segment, thereby causing the local maximum of the interlayer conductivity. Thus, the field direction that yields a local minimum of interlayer resistance indicates the short axis of the elliptic constant-energy contours of the tilted Dirac cone.

## III. TRANSPORT MEASUREMENTS ON $\alpha$-(BEDT-TTF)$_2$I$_3$

We performed the magnetotransport measurement in $\alpha$-(BEDT-TTF)$_2$I$_3$. The sample crystals were grown via the standard electrochemical method. The crystal axes were determined via X-ray diffraction (XRD). The lattice parameters obtained via XRD were in good agreement with the parameters reported in extant studies [28] (We use the same definition of the crystal axes as Ref. 28 in this paper). The electrodes were formed on the top and bottom surfaces of crystals using the gold-paste for four-terminal interlayer resistance measurements. The sample was mounted in the piston-cylinder-type pressure cell to align its orientation relative to the pressure cell, and it was set in a split-type superconducting magnet system with a rotation mechanism where the rotation origin was adjusted using the reflection of the laser beam.

The inset of Fig. 2 shows the schematic phase diagram of $\alpha$-(BEDT-TTF)$_2$I$_3$ under hydrostatic pressures. The critical temperature $T_{CO}$ of the CO phase is suppressed by the pressure, and the Dirac SM phase is stabilized to zero temperature above the critical pressure $P_c$. As mentioned above, we refer to the CO region just below $P_c$ as the weak CO state. Although the CO transition is the first-order phase transition, the weak

CO state was experimentally clarified as a massive Dirac fermion state with a small energy gap [22].

The main panel of Fig. 2 shows the temperature dependence of interlayer resistance $R_{zz}$ of α-(BEDT-TTF)$_2$I$_3$ under hydrostatic pressures. When the temperature decreases at $P$ = 1.2 GPa, the resistance begins to increase at approximately $T$ = 50 K, and this corresponds to the transition from the Dirac SM to the weak CO state. In the weak CO state, $T_{CO}$ and the resistance are significantly reduced due to the small gap. It should be noted that metallic temperature dependence ($dR_{zz}/dT > 0$) is observed in the weak CO state as indicated by a dashed circle. Around the region, a finite spin gap was observed by the NMR measurement despite the suppression of the charge gap [23]. There is no established explanation for the anomalous behaviors yet.

At higher pressures, the resistance exhibits metallic temperature dependence corresponding to the Dirac SM state with the exception of the low temperature region. At low temperatures below 5 K, the resistance exhibits a significant insulating increase ($dR_{zz}/dT < 0$) in all pressures (Dirac SM and weak CO states) as shown in Fig. 2. The insulating behavior corresponds to another unexplained problem [23]. The possibility of a small gap due to the spin-orbit interaction is discussed as a potential explanation [29, 30].

IV. MAGNETOTRANSPORT MEASUREMENTS AND IN-PLANE ANISOTROPY OF THE DIRAC CONE

The interlayer magetoresistance was measured under in-plane magnetic fields as a function of its azimuthal angle measured from the crystalline **a**-axis towards the **b**-axis. The measurements were performed above 1.2 GPa where the resistance is in the measurable range at low temperatures in the CO phase. Even a small misalignment in the sample crystal can lead to the superposition of the normal field effect, which yields sharp peak structures reflecting negative interlayer magnetoresistance [14, 15]. To prevent mixing of the normal field effect without any reproducibility, we performed measurements at sufficiently low magnetic fields where the normal field effect disappears. Figure 3(a) shows the angle-dependence of interlayer resistance at a low temperature (4 K) and a weak magnetic field (0.1 T) for several pressures between 1.2 and 2.0 GPa. The resistance exhibits a sinusoidal change, and it exhibits local minima at approximately 60° and 240°. This suggests that the short axis of elliptic constant-energy contours of the Dirac cones is tilted from the **a**-axis towards the **b**-axis by approximately 60°, and thus the long axis is oriented at approximately −30°.

It should be noted that low pressure data ($P < 1.5$ GPa) was considered in the weak CO phase with a small energy gap at the Dirac point. The quasi-particles activated thermally beyond the small CO gap must display angle-dependence corresponding to the elliptic constant-energy contours of gapped Dirac cones. The minimum angle exhibits weak pressure dependence as shown in Fig. 3(b). The minimum angle increases from 50° to 60° with increases in the pressure from 1.5 to 2.0 GPa while it decreases below 1.5 GPa. Thus, in approximate terms, the direction of the Dirac cone axis appears to exhibit opposite changes in the Dirac SM and weak CO states with respect to pressure.

V. TEMPERATURE-INDUCED CHANGE OF ANGLE-DEPENDENT PATTERN

The angle-dependent pattern of interlayer resistance exhibits unexpected temperature dependence. Figure 4(a) shows the angle-dependent pattern in the Dirac SM state ($P = 2.0$ GPa) at several temperatures. When the temperature increases from 4.0 K, the minimum angle rapidly switches from approximately 60° to 0° at approximately $T = 30$–40 K. The same feature is observed in the weak CO state ($P = 1.2$ GPa) as shown in Fig. 4(b). With respect to various pressures between 1.2 and 2.0 GPa, the temperature dependences of the minimum angle are plotted in Fig. 4(c). As mentioned above, the minimum angle is in the range of 50–60° at low temperatures both in the Dirac SM and weak CO states although it is slightly dependent on pressure. When the temperature increases, the minimum angle decreases rapidly at approximately 30–40 K both in the Dirac SM and weak CO states and reaches almost 0° (parallel to the **a**-axis) at higher temperatures. It is important to note that the behavior is independent of pressure. This suggests that the behavior mainly originates from the change in the thermal distribution and not the change in electronic structure. The distribution effect is also supported by the fact that the resistance does not exhibit a characteristic structure at approximately 30–40 K as shown in Fig. 2.

The change in the minimum angle at approximately 30–40 K is explained by the thermal excitation on the van Hove singularity. We assume that the ridges of two Dirac cones meet at a saddle point of the 2D dispersion of the conduction or valence band as shown schematically in Fig. 1(c). The saddle point must correspond to one of the symmetric points in the 2D Brillouin zone due to time reversal symmetry. It yields a van Hove singularity with a divergent peak of density of states (DOS). At low temperatures, electrons and holes are distributed only around the Dirac point where the DOS is low. When the temperature increases, electrons or holes are thermally excited on the van Hove singularity. The van Hove singularity exhibits high DOS, and thus the contribution of the excited carriers dominates the resistance anisotropy at high temperatures. Therefore, the observed change in the resistance minimum angle must correspond to the excitation from the Dirac point to the van Hove singularity, and the minimum angle at higher temperatures reflects the anisotropy of the saddle point.

If the aforementioned picture is accurate, then this implies that a van Hove singularity exists approximately 3–4 meV above or below the Dirac point, and the value is not extremely sensitive to pressure. The minimum angle at high temperatures (approximately 0°) suggests that most parts of hyperbolic constant-energy contours around the saddle point are normal with respect to the crystalline **a**-axis, and thus the ridgeline of the saddle point (median line of two asymptotes of hyperbolic constant-energy contours) is parallel to the **b**-axis.

The thermal excitation onto the van Hove singularity was already observed as the shoulder structure of local spin susceptibility measured by NMR [20]. Although the present excitation energy (3–4 meV) is slightly lower than the NMR estimation (6–12 meV), the value is in the same order. The value still exceeds the excitation energy from the ground Landau level (zero mode) to the first Landau level, which is less than 2 meV below 10 T [16]. Therefore, the present result does not conflict with those obtained in extant experimental studies.

The configurations of the Dirac cone and the van Hove singularity suggested by the present study (which is independent of any band models) disagree with those of conventional band models [3-13]. Specifically, they predict that the long axis of Dirac cone ellipse and ridgeline of the van Hove singularity are typically tilted from **a**- to **b**-axis by 70° and 30°, respectively. The reason for the inconsistency is unclear at the present stage. Hence, further investigations are potentially necessary.

## VI. SUMMARY

In conclusion, we developed an experimental method to study the anisotropy of electronic structure of layered conductors based on the dependence of interlayer magnetoresistance on the azimuthal angle of in-plane magnetic fields. We applied the method to an organic Dirac fermion system $\alpha$-(BEDT-TTF)$_2$I$_3$. The long axis of elliptic constant-energy contours of the Dirac cone is tilted by approximately $-30°$ from **a**- to **b**-axis. The configuration slightly depend on the pressure. In addition, the existence of a van Hove singularity is strongly suggested at 30–40 K above or below the Dirac point. The ridgeline of saddle point dispersion of van Hove singularity is almost parallel to **b**-axis. Further examination might be needed since the results disagree with the conventional band models.

## ACKNOWLEDGEMENTS

The authors thank Prof. Woun Kang and Prof. Y. Suzumura for their insightful discussions and valuable comments. The X-ray diffraction measurement was performed using the facility at the X-ray laboratory, Institute for Solid State Physics, University of Tokyo. The study was supported by JSPS KAKENHI Grant Numbers JP25107003 and JP16H03999.

*corresponding author, osada@issp.u-tokyo.ac.jp

#present address: *International Center for Material Nanoarchitectonics, National Institute for Materials Science, 1-1 Namiki, Tsukuba, Ibaraki 305-0044, Japan.*

**Figure 1** (Mori *et al.*)

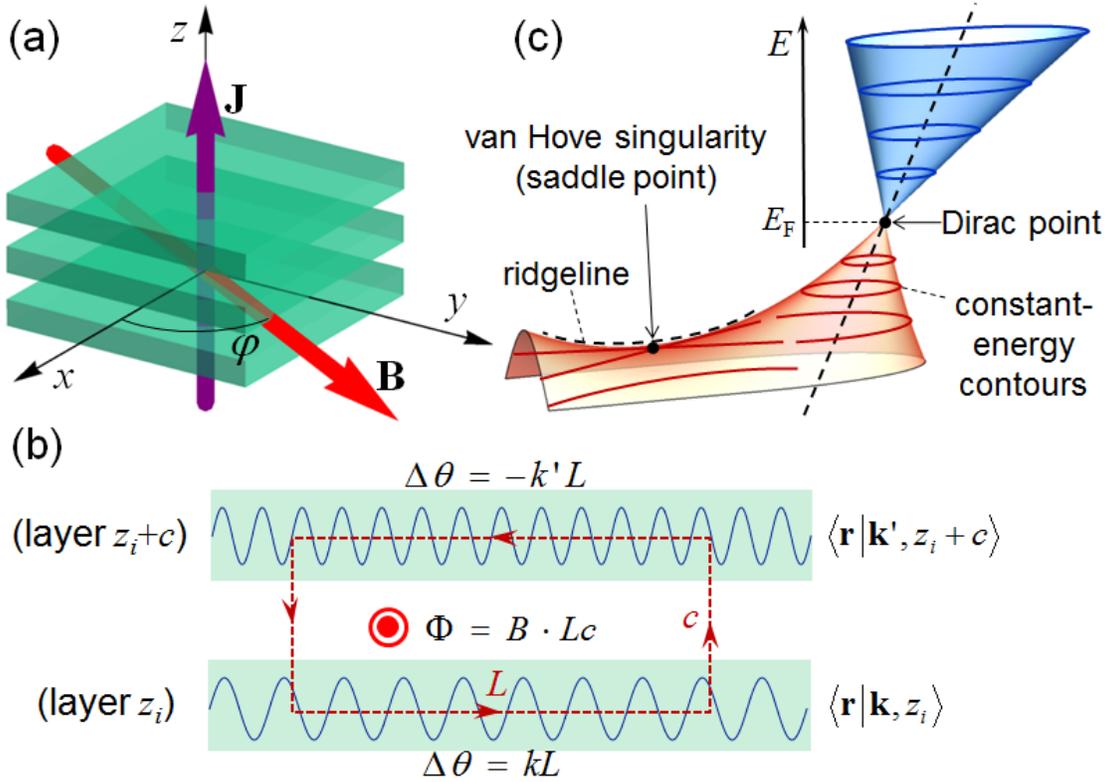

**FIG. 1.** (color online)

(a) Schematic view of the experimental configuration. Interlayer resistance is measured as a function of the azimuthal angle $\varphi$ of in-plane magnetic field. (b) Wave number shift between two wave functions coupled by interlayer tunneling. The dashed line indicates the loop that is arbitrarily selected. (c) Schematic band dispersion of the 2D Dirac fermion system. An example of the saddle point exhibiting van Hove singularity is also shown.

**Figure 2** (Mori *et al.*)

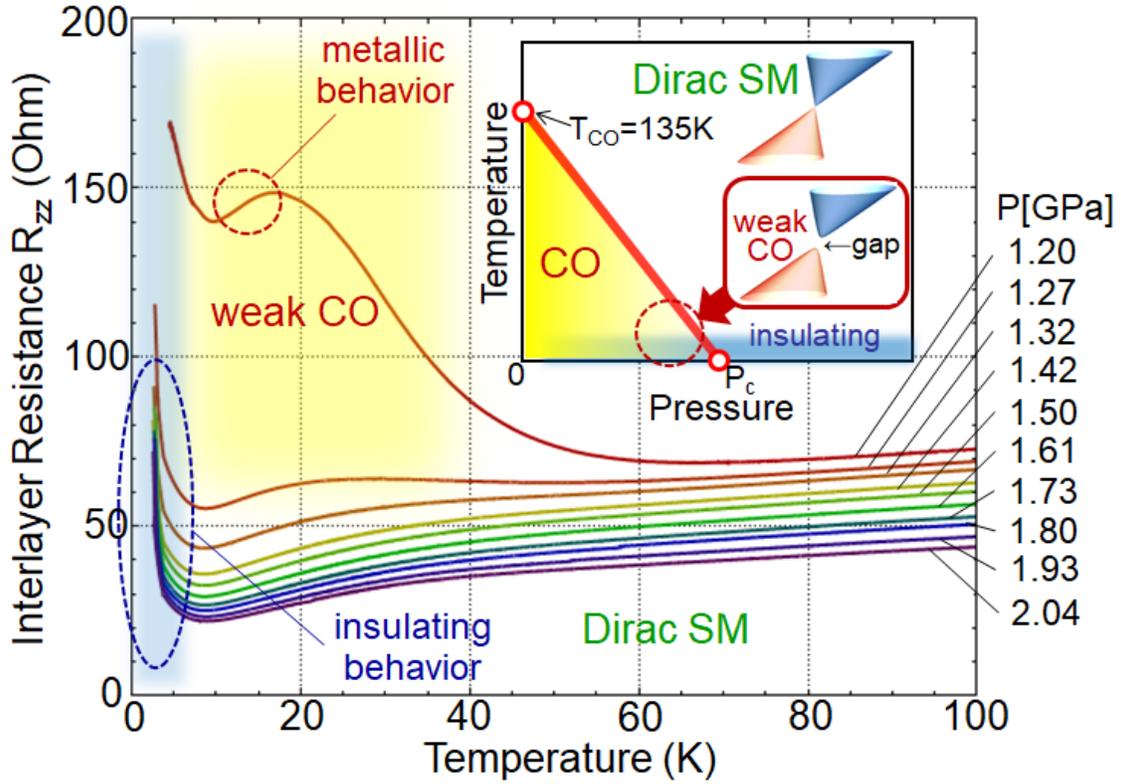

**FIG. 2.** (color online)

Temperature dependence of interlayer resistance of α-(BEDT-TTF)$_2$I$_3$ under several hydrostatic pressures at zero magnetic field. The inset shows a schematic phase diagram. The high-pressure region just below the critical pressure $P_c$ in the CO phase is referred as the weak CO state. It is considered to exhibit a gapped Dirac cone dispersion. The metallic behavior in the weak CO state is indicated by a dashed circle. Unclarified insulating behavior is observed at low temperatures in the Dirac semimetal phase as denoted by a dashed oval.

**Figure 3** (Mori *et al.*)

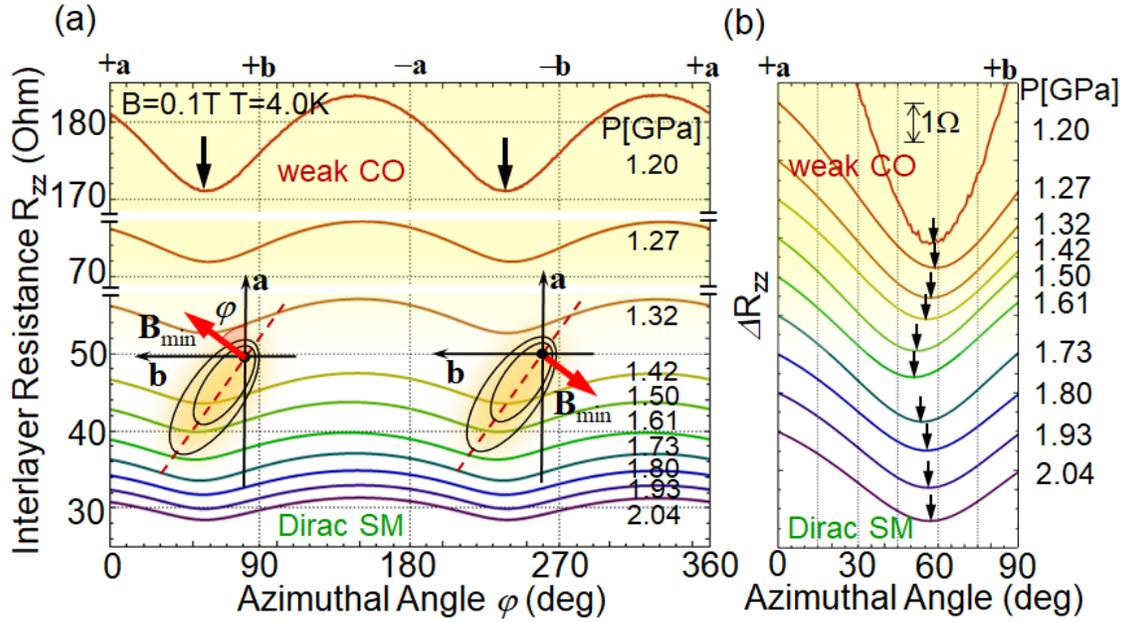

**FIG. 3.** (color online)

Interlayer magnetoresistance of α-(BEDT-TTF)$_2$I$_3$ as a function of the azimuthal angle of the in-plane magnetic field under several hydrostatic pressures. The field strength is fixed at 0.1 T, and the temperature is 4.0 K. (a) Whole angle dependence with two-fold symmetry. Two resistance minima indicated by arrows correspond to the short axis directions of elliptic constant-energy contours of Dirac cones as indicated in insets. (b) Angle dependence in an enlarged scale. The resistance minimum exhibits weak pressure dependence.

**Figure 4** (Mori *et al.*)

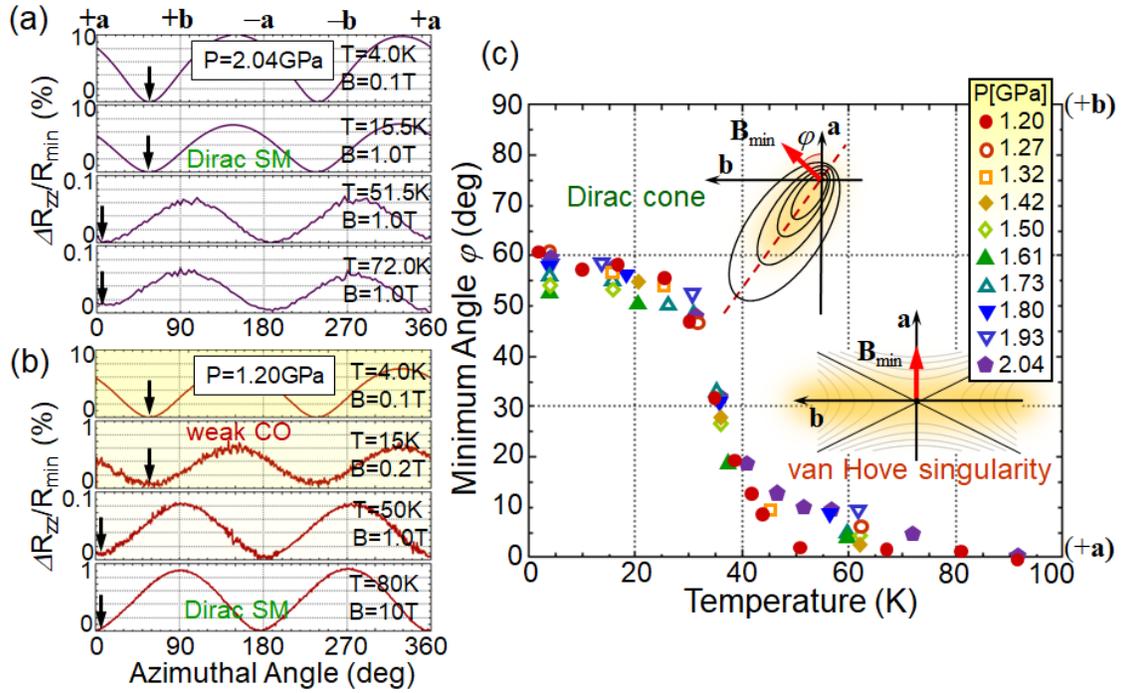

**FIG. 4.** (color online)

Interlayer magnetoresistance of $\alpha$-(BEDT-TTF)$_2$I$_3$ as a function of the azimuthal angle of the in-plane magnetic field at several temperatures. The field strength was selected to obtain clear angular dependence. (a) Angle dependence at 2.0 GPa (Dirac SM phase). (b) Angle dependence at 1.2 GPa (weak CO state at low temperatures). (c) Temperature dependence of the resistance minimum angle for different pressures. As shown in the insets, the minimum angle at low temperatures (50–60°) and high temperatures (0–10°.) reflect the orientation of the Dirac cone and van Hove singularity, respectively.

19